\documentclass[eqsecnum,aps,showpacs]{revtex4}

\def\ssqr#1#2{{\vbox{\hrule height #2pt
      \hbox{\vrule width #2pt height#1pt \kern#1pt\vrule width #2pt}
      \hrule height #2pt}\kern- #2pt}}

\def\bsqr{\ssqr{15}{.3}}

\def\nboxS{\vbox{$\overbrace{\hbox{$\bsqr\bsqr\bsqr\raise5.0pt\hbox{$\,\cdot
\cdot\cdot\,$}\bsqr\bsqr$}}^{\displaystyle N_c}$}}

\def\nboxMS{\vbox{\hbox{$\bsqr\!\overbrace{\bsqr\bsqr\raise5.0pt\hbox{$\,\cdot
\cdot\cdot\,$}\bsqr\bsqr}^{\displaystyle N_c \! - \! 2}$}\nointerlineskip 
\kern-.2pt\hbox{$\bsqr$}}}

\def\nboxA{\vbox{\hbox{$\bsqr\!\overbrace{\bsqr\bsqr\raise5.0pt\hbox{$\,\cdot
\cdot\cdot\,$}\bsqr\bsqr}^{\displaystyle N_c \! - \! 3}$}\nointerlineskip 
\kern-.2pt\hbox{$\bsqr$}\nointerlineskip
\kern-.2pt\hbox{$\bsqr$}}}

\def\tm{\tilde{m}}

\begin{document}

\begin{flushright}
DOE-ER-40762-285\\
UMD PP\#03-056
\end{flushright}

\count255=\time\divide\count255 by 60 \xdef\hourmin{\number\count255}
  \multiply\count255 by-60\advance\count255 by\time
 \xdef\hourmin{\hourmin:\ifnum\count255<10 0\fi\the\count255}

\newcommand{\xbf}[1]{\mbox{\boldmath $ #1 $}}

\newcommand{\sixj}[6]{\mbox{$\left\{ \begin{array}{ccc} {#1} & {#2} &
{#3} \\ {#4} & {#5} & {#6} \end{array} \right\}$}}

\newcommand{\threej}[6]{\mbox{$\left( \begin{array}{ccc} {#1} & {#2} &
{#3} \\ {#4} & {#5} & {#6} \end{array} \right)$}}

\title{Compatibility of Quark and Resonant Picture Excited Baryon
Multiplets in $1/N_c$}

\author{Thomas D. Cohen}
\email{cohen@physics.umd.edu}

\affiliation{Department of Physics, University of Maryland, College
Park, MD 20742-4111}

\author{Richard F. Lebed}
\email{Richard.Lebed@asu.edu}

\affiliation{Department of Physics and Astronomy, Arizona State
University, Tempe, AZ 85287-1504}

\date{June, 2003}

\bigskip

\begin{abstract}
We demonstrate that the two major complementary pictures of large
$N_c$ baryon resonances---as single-quark orbital excitations about a
closed-shell core [giving SU(2$N_F$)$\times$O(3) multiplets], and as
resonances in meson-baryon scattering amplitudes---are completely
compatible in a specific sense: Both pictures give rise to a set of
multiplets of degenerate states, for which any complete spin-flavor
multiplet within one picture fills the quantum numbers of complete
multiplets in the other picture.  This result is demonstrated by: (i)
straightforward computation of the lowest multiplets in both pictures;
(ii) a study of the nature of quark excitations in a hedgehog picture;
(iii) direct group-theoretical comparison of the constraints in the
two pictures.
\end{abstract}


\maketitle
\thispagestyle{empty}

\newpage
\setcounter{page}{1}

\section{Introduction}\label{sec:intro}

The quark model has historically served as the paradigm for
understanding the general patterns of excited baryonic states.  In
important ways, however, this picture is unsatisfactory.  In
particular, it is entirely unclear how to obtain the quark model as a
direct consequence of fundamental QCD interactions.  At a more mundane
level, while the quark model describes only bound states of an
unchanging number of confined quarks, all of the excited baryons
observed in nature appear as resonances in scattering experiments,
which are not comprehensible without an understanding of quark
production and annihilation dynamics.  Thus, in order to compare quark
model predictions to experimental data one must make additional {\it
ad hoc\/} assumptions about the strength and mechanism of such quark
processes.

Recently, we argued~\cite{us} that large $N_c$ QCD provides insights
into both of these issues.  In particular, we showed that the
``contracted'' spin-flavor symmetry~\cite{GS,DM,Jenk,DJM1} known to
emerge for ground-state baryons directly from large $N_c$ QCD imposes
important constraints on the resonance positions (masses and widths)
of excited baryons.  These constraints imply that various resonances
with distinct $I$ and $J$ quantum numbers become degenerate as $N_c$
becomes large.  On the other hand, treatments using a large $N_c$
quark picture exhibit precisely the same degeneracy
patterns~\cite{us,PY,PS}, and thus capture at least some nontrivial
QCD dynamics.

Specifically, in Ref.~\cite{us} we demonstrated these degeneracies for
the case of the mixed-symmetry $\ell \! = \! 1$ negative-parity states
in the simple quark-shell model, where $\ell$ denotes the orbital
angular momentum of a single quark excited with respect to a ``core''
of $N_c \! - \!  1$ quarks symmetrized in spin$\times$flavor.
Explicit diagonalization of the large $N_c$ quark-picture Hamiltonian
for these states~\cite{CCGL1,CCGL2} manifests these mass degeneracies.
The purpose of the present paper is to show that the previous result
is in fact general and applies to excited states of {\em any\/}
angular momentum, parity, or symmetry.  For simplicity and clarity of
presentation, we restrict our attention to the nonstrange sector.

The {\em operator method\/} of constructing the baryon Hamiltonian in
the large $N_c$ quark-shell picture ({\it i.e.}, as a linear
combination of operators constructed from the spin and flavor degrees
of freedom of the individual quark interpolating fields~\cite{BL1})
was originally applied to the ground-state spin-flavor multiplet
containing $N$ and $\Delta$ [the large $N_c$ analogue to the SU(6)
symmetric {\bf 56}-plet]~\cite{DJM2,JL,LMR,CGO}.  References to a
number of other applications to the ground-state band are listed in
Ref.~\cite{BHL}.  Subsequent work extended the method to excited
baryons, in particular the $\ell \! = \! 1$ multiplet [the large $N_c$
analogue of the SU(6) mixed-symmetry {\bf
70}-plet]~\cite{CGKM,Goity,CCGL1,CCGL2,CC1,GSS1}, and even higher
multiplets~\cite{CC2,GSS2}.

On the other hand, the {\em consistency condition method\/},
originating directly from the contracted spin-flavor symmetry, tends
to be less represented in the literature, even though the method is
obtained directly from considerations of meson-baryon scattering
processes and thus is closely intertwined with the phenomenon of
baryon resonances.  The first such work on excited baryons~\cite{PY}
studied resonances by deriving quark-picture operators that satisfy
the contracted spin-flavor symmetry, and then used the operators to
obtain a Hamiltonian, from which matrix elements describing the
spectrum and decays of these states were obtained.

One may, however, question the validity of the operator picture in
describing intrinsically unstable systems such as baryon resonances
within a Hamiltonian formalism, since the states in the bra and ket of
its matrix elements must be assumed to exist for a sufficiently long
time in order to be described as eigenstates of a Hamiltonian (rather
than simply as bumps in a scattering amplitude).  For example, generic
large $N_c$ counting shows that meson-baryon scattering amplitudes
scale as $N_c^0$, suggesting that true baryon resonances have masses
above those in the ground-state band [which are $O(N_c^1)$] by an
amount of $O(N_c^0)$, as well as widths that also scale as $N_c^0$.
On the other hand, Refs.~\cite{PY} suggest that the $\ell \! = \! 1$
excited baryons can have widths of $O(1/N_c)$.  Whether there exist
baryon resonances exhibiting this property (and therefore rendering
them stable in the large $N_c$ limit) remains an open question.

In this paper, however, we adopt the starting point that all baryon
resonances arise from generic poles in meson-baryon scattering
amplitudes with energies above the ground state of $O(N_c^0)$ and
widths of $O(N_c^0)$.  Work along these
lines~\cite{HEHW,MP,MM,Mat3,MK} long predates the operator method
papers, and was originally derived not from a quark picture, but from
a chiral soliton picture.  In the large $N_c$ limit it was shown that
the results of this approach may be derived from a simple
prescription~\cite{MM}: In treating the scattering as a $t$-channel
process, the isospin exchanged equals the angular momentum exchanged,
the so-called $I_t \!  = \! J_t$ rule.  In Ref.~\cite{us}, this rule
was shown to be equivalent to the scattering consistency conditions
derived from the contracted spin-flavor symmetry, which again is a
consequence of large $N_c$ without additional model assumptions.

One therefore expects that, if the operator method is physically
valid, it must generate a spectrum of states compatible with the
resonances obtained from the $I_t \!  = \!  J_t$ rule.  It turns out
that such resonance multiplets may be labeled by nonnegative integers
$K$, which correspond to the ``grand spin'' of the Skyrme and other
chiral soliton models.  The ``compatibility'' of this paper's title
refers to the fact~\cite{us} that the spectrum of resonant states
obtained in the quark-shell model mixed-symmetry $\ell \! = \! 1$
multiplet consists of one member, for each allowed $I$ and $J$ quantum
number, from precisely three complete degenerate multiplets (labeled
$K \! = \! 0$, 1, and 2) in the scattering ``resonance'' picture.  In
other words, in that multiplet the (complex) mass parameter for each
one of the 13 distinct states equals either $m_{K=0}$, $m_{K=1}$, or
$m_{K=2}$.  This remarkable correspondence between two very different
physical pictures is the motivation for the present work.  As
suggested above, we show that each (nonstrange) quark-picture
multiplet is filled by a collection of resonance-picture $K$
multiplets, one mass (and width) value from each resonance-picture $K$
multiplet appearing for each allowed value of $I$ and $J$ in the
quark-picture multiplet.

This paper is arranged as follows: In Sec.~\ref{resonance}, we discuss
in greater detail the resonance picture, the significance of $K$ spin,
and linear relations between $S$-matrix amplitudes.
Section~\ref{quark} lays out all relevant baryon spin-flavor
multiplets for arbitrary $N_c$, their decomposition into separate spin
and flavor multiplets, and relations between quantum numbers of their
nonstrange members.  In Sec.~\ref{compat} we exhibit explicitly the
compatibility of quark-shell and resonance picture states.
Section~\ref{gen} explains why this compatibility occurs, first using
a general description in terms of Skyrme model hedgehog
configurations, and second using explicit group-theoretical counting.
Discussion and conclusions appear in Sec.~\ref{discuss}.

\section{Meson-Nucleon Scattering  Picture} \label{resonance}

We begin with a brief review of the origin of degeneracies among
baryonic resonances.  The key to the analysis is the existence of
linear relations among the $S$ matrices of various channels in
meson-nucleon scattering (or more generally, scattering of mesons off
ground-state band baryons)~\cite{HEHW,MP,MM,Mat3,MK}.  Here we need
only study the scattering of $\pi$ or $\eta$ mesons off a ground-state
band baryon; these cases are sufficient to connect to all quantum
numbers of interest.  Note that in the physical world the notion of
describing the scattering of unstable mesons such as $\eta$ off
baryons is problematic.  However, it is sensible to talk about such
scattering in large $N_c$ since these mesons are stable in the large
$N_c$ world.  The $S$ matrices are related by
\begin{eqnarray}
S_{LL^\prime R R^\prime IJ}^\pi & = &\sum_K (-1)^{R^\prime - R}
\sqrt{(2R+1)(2R^\prime+1)} (2K+1) \left\{ \begin{array}{ccc} K &
I & J\\ R^\prime & L^\prime & 1 \end{array} \right\} \left\{
\begin{array}{ccc} K & I & J \\ R & L & 1 \end{array} \right\}
s_{KL^\prime L}^\pi , \nonumber \\
S_{L R J}^\eta & = & \sum_K
\delta_{KL} \, \delta (LRJ) \, s_{K}^\eta .
\label{MPeqn}
\end{eqnarray}
The notation is as follows: In the case of $\pi$ scattering, the
incoming baryon spin (which equals its isospin for the nonstrange
members of the ground-state band) is denoted as $R$, final baryon spin
(isospin) is denoted $R^\prime$; the incident (final) $\pi$ is in a
partial wave of orbital angular momentum $L$ ($L^\prime$), and $I$ and
$J$ represent the (conserved) total isospin and angular momentum,
respectively, of the initial and final states.  $S_{LL^\prime R
R^\prime IJ}^{\pi}$ is the (isospin- and angular momentum-reduced) $S$
matrix for this channel reduced in the sense of the Wigner-Eckart
theorem, the factors in braces are $6j$ coefficients, and $s_{K
L^\prime L}^{\pi}$ are universal amplitudes that are independent of
$I$, $J$, $R$, and $R'$.  In the case of $\eta$ scattering, the fact
that the meson has $I \! = \! 0$ more tightly constrains many of the
quantum numbers.  The isospin (= spin) $R$ of the baryon is unchanged
and moreover equals the total isospin $I$ of the intermediate state.
The orbital angular momentum $L$ of the $\eta$ remain unchanged in the
process due to the $I_t=J_t$ rule, and $J$ denotes the total
angular momentum of the state, which is constrained by the triangle
rule $\delta (L R J)$.  $S_{L R J}^{\eta}$ is the reduced scattering
amplitude and $s_{K}^{\eta}$ are universal amplitudes independent of
$J$.  The linear relations among the scattering amplitudes can be seen
from the structure of Eqs.~(\ref{MPeqn}).  The key point is simply
that there are more $S_{LL^\prime R R^\prime IJ}^{\pi}$ amplitudes
than there are $s_{K L^\prime L}^{\pi}$ amplitudes. Thus at leading
order in $1/N_c$, there are linear constraints between the
$S_{LL^\prime R R^\prime IJ}^{\pi}$ amplitudes.  There are also
more $S_{L R J}^{\eta}$ amplitudes than $s_{K}^\eta$ amplitudes,
yielding more nontrivial relations.

Equations~(\ref{MPeqn}) were first derived in the context of the
Skyrme model~\cite{HEHW,MP,MK,MM,Mat3}.  In this picture, as in other
chiral soliton models, the soliton at the classical or mean-field
level (which dominates as $N_c \rightarrow \infty)$ breaks both the
rotational and isospin symmetries but is invariant under ${\bf K}
\equiv {\bf I}+ {\bf J}$.  Accordingly, the intrinsic dynamics of the
soliton commutes with the ``grand spin'' of ${\bf I} + {\bf J}$, and
excitations can be labeled by $K$.  This is the $K$ of
Eqs.~(\ref{MPeqn}).  Note that the physical states are projected from
the hedgehogs, so that ${\bf K}$ of the physical state is {\em not\/}
just ${\bf I} + {\bf J}$, but rather represents the grand spin of the
underlying intrinsic state.

The derivation of Eqs.~(\ref{MPeqn}) from the Skyrme model has the
advantage of suggesting a clear physical picture in which the $K$
quantum number has a simple interpretation.  Of course, it has the
disadvantage of being based on a model rather than directly on large
$N_c$ QCD.  However, these relations are, in fact, exact results in
large $N_c$ QCD and do not depend on any additional model assumptions.
A direct derivation based on large $N_c$ consistency rules~\cite{DJM1}
and exploiting the famous $I_t \! = \! J_t$ rule~\cite{MM} is given in
the appendix of Ref.~\cite{us}.

Equations~(\ref{MPeqn}) can be obtained from a more general
relation~\cite{MM}---again equivalent to the $I_t \! = \! J_t$
rule---describing meson-baryon scattering of nonstrange particles
\begin{equation}
m + B \to m^\prime + B^\prime ,
\end{equation}
where $m \, (m^\prime)$ is a meson of spin $s \, (s^\prime)$ and
isospin $i \, (i^\prime)$, $B \, (B^\prime)$ is a baryon in the
ground-state multiplet with spin=isospin $R \, (R^\prime)$, and the
total spin angular momentum (not including relative orbital angular
momentum) of the meson and baryon is denoted $S \, (S^\prime)$.
Adopting the remainder of the symbols from Eqs.~(\ref{MPeqn}) and
abbreviating the multiplicity $2X\!+\!1$ of an SU(2) representation of
quantum number $X$ by $[X]$, then Eq.~(3) of Ref.~\cite{MM} reads
\begin{eqnarray}
S_{L L^\prime S S^\prime I J} & = & \sum_{K, \tilde{K} ,
\tilde{K}^\prime} [K]
([R][R^\prime][S][S^\prime][\tilde{K}][\tilde{K}^\prime])^{1/2}
\nonumber \\
& & \times \left\{ \begin{array}{ccc}
L & i & \tilde{K} \\
S & R & s \\
J & I & K \end{array} \right\}
\left\{ \begin{array}{ccc}
L^\prime & i^\prime & \tilde{K}^\prime \\
S^\prime & R^\prime & s^\prime \\
J & I & K \end{array} \right\}
\tau_{K \tilde{K} \tilde{K}^\prime L L^\prime} . \label{Mmaster}
\end{eqnarray}
In the cases considered here, the mesons are both spinless ($s \!  =
\! s^\prime \!= \! 0$), which implies the collapse of the $9j$ symbols
to $6j$ symbols, and forces $S \to R$, $S^\prime \to R^\prime$, and
$\tilde{K} \! = \! \tilde{K}^\prime \! = \! K$.  The first of
Eqs.~(\ref{MPeqn}) is then obtained by taking $i \! = \! i^\prime \! =
\! 1$ and $s^\pi_{KLL^\prime} = (-1)^{L-L^\prime}
\tau_{KKKLL^\prime}$, while the second of Eqs.~(\ref{MPeqn}) is
obtained by taking $i \! = \!  i^\prime \! = \! 0$ [which further
collapses the $6j$ symbols, leaving only $\delta_{LL^\prime} \,
\delta_{RR^\prime} \, \delta_{IR} \, \delta(LRJ)$] and by taking
$s^\eta_K = \tau_{KKKLL}$.

The pattern of degeneracies of the masses and widths of resonances
falls out immediately from the structure of Eqs.~(\ref{MPeqn}),
provided one defines the resonance position to be at the pole.  Since
resonances are broad, there is some ambiguity in defining precisely
what one means by the mass of the resonance.  Moreover, since
generically baryon resonances in the large $N_c$ limit need not be
narrow, these ambiguities can be large.  Perhaps the cleanest
theoretical way to define the resonance position is via analytic
continuation of the scattering amplitude in the complex energy plane
(with three-momentum fixed to zero) and to define a resonance as
occurring at the point at which the amplitude develops a pole. The
resonance mass and width are then the real and imaginary parts of the
complex pole position.  Note that with this definition degeneracies
automatically follow.  In order for one of the $S_{LL^\prime R
R^\prime IJ}^{\pi}$ amplitudes to diverge, one of the $s_{KL^\prime
L}^{\pi}$ amplitudes must diverge.  However, since the $s_{KL^\prime
L}^{\pi}$ amplitudes contribute to multiple channels, all of these
channels must have resonances at the same position, which means that
the masses and widths of certain resonances in different channels must
be degenerate.  As seen in Ref.~\cite{us}, the observed pattern of
degeneracies for the low-lying negative-parity $\ell \! = \! 1$
baryons obtained for the resonances is identical to that obtained in
the large $N_c$ quark-shell picture.  Below we show that the
coincidence between the pattern of degeneracies for baryon resonances
in large $N_c$ QCD and states in the large $N_c$ quark-shell picture
holds far more generally.

\section{States in the Quark-Shell Model Picture}\label{quark}

\begin{figure}[ht]
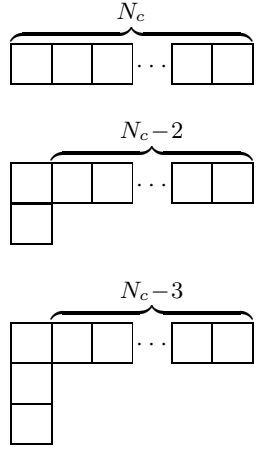

  \begin{centering}

\centerline{$$\nboxS$$}
\bigskip

\centerline{$$\nboxMS$$}
\bigskip

\centerline{$$\nboxA$$}
\bigskip

\caption{Young tableaux for the SU($2N_F$) S, MS, and A
spin-flavor representations, respectively.  For $N_F \! = \! 3$ these
are the familiar {\bf 56}, {\bf 70}, and {\bf 20}, respectively.}
\label{young}
\end{centering}
\end{figure}

The analysis of spin-flavor multiplets in the quark-shell picture is a
straightforward, albeit tedious, extension of methods familiar from
$N_c \! = \! 3$.  For $N_c \! = \! 3$ only three spin-flavor
representations occur: (i) Completely symmetric (S), which is a {\bf
56} for $N_F \! = \! 3$ or a {\bf 20} for $N_F \! = \! 2$; (ii)
mixed-symmetry (MS), which is a {\bf 70} for $N_F \! = \! 3$ or a {\bf
20} for $N_F \! = \! 2$; and (iii) completely antisymmetric (A), which
is a {\bf 20} for $N_F \! = \! 3$ or a {\bf 4} for $N_F \! = \! 2$.
For $N_c \! > \! 3$ many other spin-flavor representations are
possible, but these do not interest us since they decouple in the
physical limit $N_c \! = \! 3$.  It should be noted that the A
representation is only fully antisymmetric for $N_c \! = \! 3$, but we
maintain the label A for $N_c \! > \! 3$.  The $N_c \! > \! 3$
generalizations of the S, MS, and A representations are presented in
Fig.~\ref{young}.  One may assume (as done here) that the additional
$(N_c \! - \! 3)$ quarks in the lowest-energy multiplets appear in a
completely symmetric spin-flavor combination, and examine whether this
ansatz gives a phenomenologically viable $1/N_c$ expansion.
Alternately, although baryon spin-flavor representations with more
than one pair of antisymmetrized indices exist for $N_c \! > \! 3$,
such states do not exist in the physical $N_c \! = \! 3$ universe and
are thus ignored in the following analysis.

The first step in the analysis is to decompose SU(2$N_F$) ${\rm spin}
\! \times \! {\rm flavor}$ representations into separate SU(2) spin
and SU($N_F$) flavor representations.  This is a fairly
straightforward but tedious exercise in pairing spin and flavor
representations with the correct overall spin-flavor transformation
properties.  To illustrate the results, we employ the SU($N_F$) Dynkin
label, which is an ($N_F\!-\!1$)-plet [$n_1, n_2, \ldots, n_{N_F-1}$]
of nonnegative integers $n_r$ describing the Young diagram of the
representation: The number of boxes in row $r$ of the Young diagram
exceeds the number in row $r \!  + \!  1$ by $n_r$.  In this notation,
the S multiplet is [$N_c, 0, 0, \ldots , 0$], MS is [$N_c \! - \! 2,
1, 0, 0, \ldots , 0$], and A is [$N_c \! - \! 3, 0, 1, 0, 0, \ldots ,
0$].  SU(2) spin representations can also be represented in this way,
but it is more convenient to use the quantum number $S$ as a label.

One then finds the (flavor, $S$) representations
\begin{eqnarray}
{\rm S} \equiv [N_c, 0, 0, \ldots 0] & = &
\bigoplus_{n=0}^{(N_c-1)/2} \left( \left[ 2n+1, \frac 1 2 (N_c-1) - n,
0, 0, \ldots, 0 \right] , n + \frac 1 2 \right) , \label{Sdecomp} \\
{\rm MS} \equiv [N_c \! - \! 2, 1, 0, 0, \ldots, 0] & = &
\bigoplus_{n=0}^{(N_c-5)/2} \left( \left[ 2n+2, \frac 1 2 (N_c-5) - n,
1, 0, 0, \ldots , 0 \right] , n + \frac 1 2 \right) \nonumber \\ & &
\bigoplus_{n=0}^{(N_c-3)/2} \left( \left[ 2n, \frac 1 2 (N_c-3) - n,
1, 0, 0, \ldots , 0 \right] , n + \frac 1 2 \right) \nonumber \\ & &
\bigoplus_{n=0}^{(N_c-3)/2} \left( \left[ 2n+1, \frac 1 2 (N_c-1) -
n, 0, 0, \ldots , 0 \right] , n + \frac 3 2 \right) \nonumber \\ & &
\bigoplus_{n=0}^{(N_c-3)/2} \left( \left[ 2n+1, \frac 1 2 (N_c-1) -
n, 0, 0, \ldots , 0 \right] , n + \frac 1 2 \right) \nonumber \\ & &
\bigoplus_{n=0}^{(N_c-3)/2} \left( \left[ 2n+3, \frac 1 2 (N_c-3) -
n, 0, 0, \ldots , 0 \right] , n + \frac 1 2 \right) , \label{MSdecomp}
\\
{\rm A} \equiv [N_c \! - \! 3, 0, 1, 0, 0, \ldots , 0] & = &
\bigoplus_{n=0}^{(N_c-5)/2} \left( \left[ 2n+1, \frac 1 2 (N_c-5) - n,
0, 1, 0, 0, \ldots, 0 \right] , n + \frac 3 2 \right) \nonumber \\ & &
\bigoplus_{n=0}^{(N_c-5)/2} \left( \left[ 2n+1, \frac 1 2 (N_c-5) - n,
0, 1, 0, 0, \ldots, 0 \right] , n + \frac 1 2 \right) \nonumber \\ & &
\bigoplus_{n=0}^{(N_c-7)/2} \left( \left[ 2n+3, \frac 1 2 (N_c-7) - n,
0, 1, 0, 0, \ldots, 0 \right] , n + \frac 1 2 \right) \nonumber \\ & &
\bigoplus_{n=0}^{(N_c-5)/2} \left( \left[ 2n+2, \frac 1 2 (N_c-5) - n,
1, 0, 0, \ldots, 0 \right] , n + \frac 3 2 \right) \nonumber \\ & &
\mbox{\Large 2} \!\!\!\!\!
\bigoplus_{n=0}^{(N_c-5)/2} \left( \left[ 2n+2, \frac 1 2 (N_c-5) - n,
1, 0, 0, \ldots, 0 \right] , n + \frac 1 2 \right) \nonumber \\ & &
\bigoplus_{n=0}^{(N_c-5)/2} \left( \left[ 2n, \frac 1 2 (N_c-3) - n,
1, 0, 0, \ldots, 0 \right] , n + \frac 1 2 \right) \nonumber \\ & &
\bigoplus_{n=0}^{(N_c-7)/2} \left( \left[ 2n+4, \frac 1 2 (N_c-7) - n,
1, 0, 0, \ldots, 0 \right] , n + \frac 1 2 \right) \nonumber \\ & &
\bigoplus_{n=0}^{(N_c-7)/2} \left( \left[ 2n+1, \frac 1 2 (N_c-7) - n,
2, 0, 0, \ldots, 0 \right] , n + \frac 1 2 \right) \nonumber \\ & &
\bigoplus_{n=0}^{(N_c-5)/2} \left( \left[ 2n+1, \frac 1 2 (N_c-1) - n,
0, 0, \ldots, 0 \right] , n + \frac 3 2 \right) \nonumber \\ & &
\bigoplus_{n=0}^{(N_c-5)/2} \left( \left[ 2n+3, \frac 1 2 (N_c-3) - n,
0, 0, \ldots, 0 \right] , n + \frac 1 2 \right) \nonumber \\ & &
\bigoplus_{n=0}^{(N_c-3)/2} \left( \left[ 2n, \frac 1 2 (N_c-3) - n,
1, 0, 0, \ldots, 0 \right] , n + \frac 3 2 \right) \nonumber \\ & &
\bigoplus_{n=0}^{(N_c-3)/2} \left( \left[ 2n+1, \frac 1 2 (N_c-1) - n,
0, 0, \ldots, 0 \right] , n + \frac 1 2 \right) . \label{Adecomp}
\end{eqnarray}

For example, in the $N_F \! = \! 3$, $N_c \! = \! 3$ case [the
familiar SU(6) $\to$ SU(3)$\times$SU(2) decomposition], any sum with
an upper limit smaller than $(N_c-3)/2$ vanishes, as does any
multiplet with a nonzero value in the fourth (or higher) entry of a
flavor multiplet Dynkin symbol, since this would require four quark
flavors among which to antisymmetrize.  All that remains in this case
is the $n \! = \! 0$ and 1 members of S = {\bf 56}, which are ({\bf
8},\,1/2) and ({\bf 10},\,3/2), respectively; the $n \! = \! 0$ member
of each of the last four sums in MS = {\bf 70}, which are ({\bf
1},\,1/2), ({\bf 8},\,3/2), ({\bf 8},\,1/2), and ({\bf 10},\,1/2),
respectively; and the $n \! = \! 0$ member of each of the last two
sums in A = {\bf 20}, which are ({\bf 1},\,3/2) and ({\bf 8},\,1/2),
respectively.  

More generally, we are interested only in those states with quantum
numbers that appear for $N_c \! = \! 3$, in particular $I \! = \! 1/2$
or 3/2, since isospin is a good quantum number for any $N_c$.
Furthermore, in this work we consider for simplicity only the
nonstrange states; the strange states are more difficult only for
technical group-theoretical reasons, and are deferred to future
work~\cite{future}.

These restrictions dramatically simplify the list of multiplets that
must be considered.  The tallest column of the Young diagram (with a
number of boxes equal to the {\em position\/} of the last nonzero
entry of the Dynkin label) indicates the number of distinct quark
flavors that must be present for each state of the multiplet for the
diagram to be allowed.  For our purposes, then, one may discard any
multiplet with a nonzero entry in $n_3$, $n_4$, etc., and consider
only Young diagrams with at most two rows.  The first entry of the
Dynkin label $n_1$ then indicates the maximum number of quarks of one
flavor ($u$, $d$, etc.) that may be symmetrized in any state of the
multiplet, and the second entry indicates the number of quark pairs
antisymmetrized in flavor.  Since the nonstrange states, lying in the
top row of the weight diagram, are singly degenerate, the isospin $I$
of the multiplet equals the maximum allowed $I_3$ value; this, in
turn, is just $\frac 1 2 n_1$.  Thus, only $n_1 \le 3$ need be
included.

The multiplets of interest, using the notation $I_S$ and the shorthand
$I \! = \! \frac 1 2 \to N$, $I \!= \! \frac 3 2 \to \Delta$, then
read
\begin{eqnarray}
{\rm S}  & : & N_{1/2} \oplus \Delta_{3/2} , \label{Slist} \\
{\rm MS} & : & N_{1/2} \oplus N_{3/2} \oplus \Delta_{1/2}
\overbrace{\oplus \, \Delta_{3/2} \oplus \Delta_{5/2}}^{N_c \ge 5} ,
\label{MSlist} \\
{\rm A}  & : & N_{1/2} \overbrace{\oplus \, N_{3/2} \oplus
\Delta_{1/2} \! \oplus \Delta_{3/2}}^{N_c \ge 5}
\overbrace{\, \oplus \, \Delta_{5/2}}^{N_c \ge 7} . \label{Alist}
\end{eqnarray}  
It is interesting to note that the A representation is, for large
$N_c$, much larger than the MS representation; however, for any $N_c
\! \ge \!7$ they contain exactly the same nonstrange $I \! = \!
\frac 1 2$ and $\frac 3 2$ states (a fact to be explained in
Sec.~\ref{gen}).  Yet, even for the nonstrange states the MS and A
multiplets differ for the highest values of isospin: By comparing
Eqs.~(\ref{MSdecomp}) and (\ref{Adecomp}), one finds MS contains
states $(I,S)$ = ($\frac{N_c}{2}\!-\!1,\frac{N_c}{2}$) and
($\frac{N_c}{2}, \frac{N_c}{2}\!-\!1$) that A does not.

Another interesting fact, to be used later, is that the $N$ and
$\Delta$ states in the MS or A representations in
Eqs.~(\ref{Slist})--(\ref{Alist}) (before excited quark angular
momentum $\ell$ is added) have the curious property that their total
quark spins $S$ are precisely those obtained from vectorially adding
one unit of angular momentum to the $S$ values of the $N, \Delta$ in
the S representation before $\ell$ is added [Eq.~(\ref{Slist})]; this
fact was noted in Ref.~\cite{PY}, for the first excited band of
baryons.  We shall see why this occurs in Sec.~\ref{gen}.  Let us
label this effective angular momentum {\boldmath \mbox{$\Delta$}}, so
that each value of $S_{\rm MS/A}$ satisfying $\delta (S_{\rm S}, \,
S_{\rm MS/A}, \, \Delta \! = \! 1 )$ occurs.  Then, since the
multiplets for $\ell\!\neq\!0$ (given in Table~\ref{qtable}) are
obtained through the vector addition {\bf J} = {\bf S} + {\boldmath
\mbox{$\ell$}}, and since the same total angular momentum quantum
numbers are obtained through any order of vector addition, it follows
that the $\ell\!\neq\!0$ MS/A $J$ eigenvalues may be obtained through
vector addition of {\boldmath \mbox{$\Delta$}} to those of S.  Thus,
each value of $J_{\rm MS/A}$ satisfying $\delta (J_{\rm S}, \, J_{\rm
MS/A}, \, \Delta \! = \! 1 )$ occurs, with the same multiplicity as
given by the vector addition
\begin{eqnarray}
{\bf J}_{\rm MS/A} & = & {\bf S}_{\rm MS/A} +
\mbox{\boldmath{\mbox{$\ell$}}}_{\rm MS/A} \nonumber \\ & = &
{\bf J}_{\rm S} + \mbox{\boldmath{\mbox{$\Delta$}}} \;
({\rm effectively}) \nonumber \\ & = &
{\bf S}_{\rm S} + \mbox{\boldmath{\mbox{$\ell$}}}_{\rm S} +
\mbox{\boldmath{\mbox{$\Delta$}}} \nonumber \\ & \equiv &
{\bf S}_{\rm S} + \mbox{\boldmath{\mbox{$\ell$}}}_{\rm eff} .
\label{Seff}
\end{eqnarray}
This implies that the MS/A states of a given $\ell$ occur with the
same quantum numbers and multiplicities as appear in an effective
collection of S multiplets, each with its own value $\ell_{\rm eff}$
that assumes all values satisfying $\delta (\ell_{\rm eff} , \, \ell,
\, \Delta \! = \! 1)$.

What remains is to add an orbital angular momentum $\ell$ to the total
quark spin $S$ in order to obtain the total spin $J$ of the state.
For the conventional states in which all quarks have positive parity,
the overall parity of the baryon is that contributed by the orbital
excitation, $P \! = \! (-1)^\ell$.  However, we shall see that all of
our results are blind to $P$ (except that it must be conserved in
strong interactions), and thus one may classify quark-picture states
in Table~\ref{qtable} by their SU($2N_F$)$\times$O(3) representation
without the need of specifying $P$.

\begin{table}
\caption{Nonstrange states of $I \! = \! \frac 1 2$ and $\frac 3 2$ in
the large $N_c$ S, MS, and MA spin-flavor representations and various
values of the excited quark orbital angular momentum $\ell$.
\label{qtable}}
\medskip
\begin{tabular}{ccccl}
Representation &\mbox{  }& $\ell$ &\mbox{  }& \hspace{3em} States \\
\hline\hline
S && 0 && $N_{1/2} \oplus \Delta_{3/2}$ \\
  && 1 && $N_{1/2} \oplus N_{3/2} \oplus \Delta_{1/2} \oplus
\Delta_{3/2} \oplus \Delta_{5/2}$ \\
  && 2 && $N_{3/2} \oplus N_{5/2} \oplus \Delta_{1/2} \oplus
\Delta_{3/2} \oplus \Delta_{5/2} \oplus \Delta_{7/2}$ \\
  && 3 && $N_{5/2} \oplus N_{7/2} \oplus \Delta_{3/2} \oplus
\Delta_{5/2} \oplus \Delta_{7/2} \oplus \Delta_{9/2}$ \\
\hline
MS or A && 0 && $N_{1/2} \oplus N_{3/2} \oplus \Delta_{1/2} \oplus
\Delta_{3/2} \oplus \Delta_{5/2}$ \\
        && 1 && $2N_{1/2} \oplus 2N_{3/2} \oplus N_{5/2} \oplus
2\Delta_{1/2} \oplus 3\Delta_{3/2} \oplus 2\Delta_{5/2} \oplus
\Delta_{7/2}$ \\
        && 2 && $N_{1/2} \oplus 2N_{3/2} \oplus 2N_{5/2} \oplus N_{7/2}
\oplus 2\Delta_{1/2} \oplus 3\Delta_{3/2} \oplus 3\Delta_{5/2} \oplus
2\Delta_{7/2} \oplus \Delta_{9/2}$ \\
        && 3 && $N_{3/2} \oplus 2N_{5/2} \oplus 2N_{7/2} \oplus N_{9/2}
\oplus \Delta_{1/2} \oplus 2\Delta_{3/2} \oplus 3 \Delta_{5/2} \oplus
3\Delta_{7/2} \oplus 2\Delta_{9/2} \oplus \Delta_{11/2}$ \\
\hline\hline
\end{tabular}
\end{table}

\section{Compatibility} \label{compat}

The multiplicities of quark-picture states in
Table~\ref{qtable} may be compared immediately with those in the
meson-nucleon scattering results presented in Tables~\ref{neg} and
\ref{pos}, which in turn are obtained via Eqs.~(\ref{MPeqn}).  Simply
note first that the $K$-amplitudes appearing for a resonance of given
spin and flavor quantum numbers ({\it e.g.}, $\Delta_{1/2}$) are the
same regardless of the parity $P$: The $I_t \! = \! J_t$ rule makes no
reference to $P$.  Of course, one should not expect the poles for
$K$-amplitudes with $+$ and $-$ parities to be equal, hence the tilde
on each mass in the $P \! = \!  +$ case.  Nevertheless, this
``parity-blindness'' mirrors that found in the SU($2N_F$)$\times$O(3)
multiplets in the quark picture.

\begin{table}
\caption{Negative-parity mass eigenvalues in the quark-shell
model picture, corresponding partial waves, and their expansions in
terms of $K$-amplitudes.  The superscripts $\pi N N$, $\pi N \Delta$,
$\pi \Delta \Delta$, $\eta N N$, and $\eta \Delta \Delta$ refer to the
scattered meson and the initial and final baryons, respectively.  The
partial-wave amplitudes are derived from Eqs.~(\ref{MPeqn}).  Note
that these states are those appropriate to a large $N_c$ world (as
discussed in the text, some do not occur for $N_c \! = \! 3$).  We
only list states with total isospin 3/2 or less, and diagonal in the
scattered meson and orbital angular momentum ($L \! = \!  L^\prime$).
\label{neg}}
%
\begin{tabular}{lcccccl}
State \mbox{  } && Masses Accommodated \mbox{   } &&
\multicolumn{3}{l}{Partial Wave, $K$-Amplitudes} \\
\hline\hline
$N_{1/2}$ && $m_0$, $m_1$
   && $S^{\pi N N}_{11}$            &=& $s^\pi_{100}$ \\
&& && $S^{\eta N N}_{1 1}$          &=& $s^\eta_0$ \\
&& && $D^{\pi \Delta \Delta}_{11}$  &=& $s^\pi_{122}$ \\
\hline
$\Delta_{1/2}$ && $m_1$, $m_2$
   && $S^{\pi N N}_{31}$            &=& $s^\pi_{100}$ \\
&& && $D_{31}^{\pi \Delta \Delta}$  &=& $\frac{1}{10}
\left( s^\pi_{122} + 9 s^\pi_{222} \right)$ \\
&& && $D^{\eta \Delta \Delta}_{31}$ &=& $s^\eta_2$ \\
\hline
$N_{3/2}$ && $m_1$, $m_2$
   && $S_{13}^{\pi \Delta \Delta}$  &=& $s^\pi_{100}$ \\
&& && $D^{\pi N N}_{13}$            &=& $\frac 1 2
\left( s^\pi_{122} + s^\pi_{222} \right)$ \\
&& && $D_{13}^{\pi N \Delta}$       &=& $\frac 1 2
\left( s^\pi_{122} - s^\pi_{222} \right)$ \\
&& && $D_{13}^{\pi \Delta \Delta}$  &=& $\frac 1 2
\left( s^\pi_{122} + s^\pi_{222} \right)$ \\
&& && $D_{13}^{\eta N N}$           &=& $s^\eta_2$ \\
\hline
$\Delta_{3/2}$  && $m_0$, $m_1$, $m_2$, $m_3$
   && $S_{33}^{\pi \Delta \Delta}$  &=& $s^\pi_{100}$ \\
&& && $S_{33}^{\eta \Delta \Delta}$ &=& $s^\eta_0$ \\
&& && $D^{\pi N N}_{33}$            &=& $\frac{1}{20}
\left( s^\pi_{122} + 5 s^\pi_{222} + 14 s^\pi_{322} \right)$ \\
&& && $D^{\pi N \Delta}_{33}$ &=& $\frac{1}{5\sqrt{10}}
\left( 2 s^\pi_{122} + 5 s^\pi_{222} - 7 s^\pi_{322} \right)$ \\
&& && $D_{33}^{\pi \Delta \Delta}$  &=& $\frac{1}{25}
\left( 8 s^\pi_{122} + 10 s^\pi_{222} + 7 s^\pi_{322} \right)$ \\
&& && $D_{33}^{\eta \Delta \Delta}$ &=& $s^\eta_2$ \\
\hline
$N_{5/2}$ && $m_2$, $m_3$
   && $D^{\pi N N}_{15}$            &=& $\frac{1}{9}
\left( 2 s^\pi_{222} + 7 s^\pi_{322} \right)$ \\
&& && $D_{15}^{\pi N \Delta}$       &=& $\frac{\sqrt{14}}{9}
\left( s^\pi_{222} - s^\pi_{322} \right)$ \\
&& && $D_{15}^{\pi \Delta \Delta}$  &=& $\frac{1}{9}
\left( 7 s^\pi_{222} + 2 s^\pi_{322} \right)$ \\
&& && $D_{15}^{\eta N N}$           &=& $s^\eta_2$ \\
&& && $G_{15}^{\pi \Delta \Delta}$ &=& $s^\pi_{344}$ \\
\hline
$\Delta_{5/2}$ && $m_1$, $m_2$, $m_3$, $m_4$
   && $D^{\pi N N}_{35}$            &=& $\frac{1}{90}
\left( 27 s^\pi_{122} + 35 s^\pi_{222} + 28 s^\pi_{322} \right)$ \\
&& && $D_{35}^{\pi N\Delta}$        &=&
$\frac{1}{90} \sqrt{\frac{7}{5}}
\left( 27 s^\pi_{122} + 5 s^\pi_{222} - 32 s^\pi_{322} \right)$ \\
&& && $D_{35}^{\pi \Delta \Delta}$  &=& $\frac{1}{450}
\left( 189 s^\pi_{122} + 5 s^\pi_{222} + 256 s^\pi_{322} \right)$ \\
&& && $D_{35}^{\eta \Delta \Delta}$ &=& $s^\eta_2$ \\
&& && $G_{35}^{\pi \Delta \Delta}$ &=& $\frac 1 4 \left( s^\pi_{344}
+ 3 s^\pi_{444} \right)$ \\
&& && $G_{35}^{\eta \Delta \Delta}$ &=& $s_4^\eta$ \\
\hline
$N_{7/2}$ && $m_3$, $m_4$
   && $D_{17}^{\pi \Delta \Delta}$ &=& $s^\pi_{322}$ \\
&& && $G_{17}^{\pi N N}$ &=& $\frac{1}{12} \left( 7 s^\pi_{344} +
5 s^\pi_{444} \right)$ \\
&& && $G_{17}^{\pi N \Delta}$ &=& $\frac{\sqrt{35}}{12} \left(
s^\pi_{344} - s^\pi_{444} \right)$ \\
&& && $G_{17}^{\pi \Delta \Delta}$ &=& $\frac{1}{12} \left( 5
s^\pi_{344} + 7 s^\pi_{444} \right)$ \\
&& && $G_{17}^{\eta N N}$ &=& $s^\eta_4$ \\
\hline
$\Delta_{7/2}$ && $m_2$, $m_3$, $m_4$, $m_5$
   && $D_{37}^{\pi \Delta \Delta}$ &=& $\frac 1 5
\left( 2 s^\pi_{222} + 3 s^\pi_{322} \right)$ \\
&& && $D_{37}^{\eta \Delta \Delta}$ &=& $s^\eta_2$ \\
&& && $G_{37}^{\pi N N}$ &=& $\frac{1}{72} \left( 35 s^\pi_{344} + 33
s^\pi_{444} + 22 s^\pi_{544} \right)$ \\
&& && $G_{37}^{\pi N \Delta}$ &=& $\frac{1}{45} \sqrt{\frac 7 2}
\left( 5 s^\pi_{344} + 6 s^\pi_{444} - 11 s^\pi_{544} \right)$ \\
&& && $G_{37}^{\pi \Delta \Delta}$ &=& $\frac{1}{225} \left( 100
s^\pi_{344} + 48 s^\pi_{444} + 77 s^\pi_{544} \right)$ \\
&& && $G_{37}^{\eta \Delta \Delta}$ &=& $s^\eta_4$ \\
\hline
$N_{9/2}$ && $m_4$, $m_5$
   && $G_{19}^{\pi N N}$ &=& $\frac{1}{15} \left( 4 s^\pi_{444} + 11
s^\pi_{544} \right)$ \\
&& && $G_{19}^{\pi N \Delta}$ &=& $\frac{2}{15} \sqrt{11} \left( 
s^\pi_{444} - s^\pi_{544} \right)$ \\
&& && $G_{19}^{\pi \Delta \Delta}$ &=& $\frac{1}{15} \left( 11
s^\pi_{444} + 4 s^\pi_{544} \right)$ \\
&& && $G_{19}^{\eta N N}$ &=& $s^\eta_4$ \\
&& && $I_{19}^{\pi \Delta \Delta}$ &=& $s^\pi_{566}$ \\
\hline
\end{tabular}
\end{table}
\begin{table}
\begin{tabular}{lcccccl}
State \mbox{  } && Masses Accommodated \mbox{   } &&
\multicolumn{3}{l}{Partial Wave, $K$-Amplitudes} \\
\hline\hline
$\Delta_{9/2}$ && $m_3$, $m_4$, $m_5$, $m_6$
   && $G_{39}^{\pi N N}$ &=& $\frac{1}{90} \left( 35 s^\pi_{344} + 33
s^\pi_{444} + 22 s^\pi_{544} \right)$ \\
&& && $G_{39}^{\pi N \Delta}$ &=& $\frac{1}{90} \sqrt{\frac{11}{10}}
\left( 35 s^\pi_{344} - 3 s^\pi_{444} - 32 s^\pi_{544} \right)$ \\
&& && $G_{39}^{\pi \Delta \Delta}$ &=& $\frac{1}{900} \left(
385 s^\pi_{344} + 3 s^\pi_{444} + 512 s^\pi_{544} \right)$ \\
&& && $G_{39}^{\eta \Delta \Delta}$ &=& $s^\eta_4$ \\
&& && $I_{39}^{\pi \Delta \Delta}$ &=& $\frac{1}{10} \left( 3
s^\pi_{566} + 7 s^\pi_{666} \right)$ \\
&& && $I_{39}^{\eta \Delta \Delta}$ &=& $s^\eta_6$ \\
\hline
$\Delta_{11/2}$ && $m_4$, $m_5$, $m_6$, $m_7$
   && $G_{3,11}^{\pi \Delta \Delta}$ &=& $\frac{1}{25} \left( 12
s^\pi_{444} + 13 s^\pi_{544} \right)$ \\
&& && $G_{3,11}^{\eta \Delta \Delta}$ &=& $s^\eta_4$ \\
&& && $I_{3,11}^{\pi N N}$ &=& $\frac{1}{468} \left( 55 s^\pi_{566} +
143 s^\pi_{666} + 270 s^\pi_{766} \right)$ \\
&& && $I_{3,11}^{\pi N \Delta}$ &=& $\frac{1}{117}
\sqrt{\frac{55}{14}} \left( 14 s^\pi_{566} + 13 s^\pi_{666} - 27
s^\pi_{766} \right)$ \\
&& && $I_{3,11}^{\pi \Delta \Delta}$ &=& $\frac{1}{819} \left( 392
s^\pi_{566} + 130 s^\pi_{666} + 297 s^\pi_{766} \right)$ \\
&& && $I_{3,11}^{\eta \Delta \Delta}$ &=& $s^\eta_6$ \\
\hline\hline
\end{tabular}
\end{table}

\setcounter{table}{2}
\begin{table}
\caption{The positive-parity amplitudes.  The notation is as in
Table~\ref{neg}.\label{pos}}
\medskip
\begin{tabular}{lcccccl}
State \mbox{ } && Masses Accommodated \mbox{ } &&
\multicolumn{3}{l}{Partial Wave, $K$-Amplitudes} \\
\hline\hline
$N_{1/2}$ && $\tm_0$, $\tm_1$
   && $P^{\pi N N}_{11}$ &=& $\frac 1 3 \left( s^\pi_{011} + 2
s^\pi_{111} \right)$ \\
&& && $P^{\pi N \Delta}_{11}$ &=& $\frac{\sqrt{2}}{3} \left(
s^\pi_{011} - s^\pi_{111} \right)$ \\
&& && $P^{\pi \Delta \Delta}_{11}$ &=& $\frac 1 3 \left( 2 s^\pi_{011}
+ s^\pi_{111} \right)$ \\
&& && $P^{\eta N N}_{11}$ &=& $s^\eta_1$ \\
\hline
$\Delta_{1/2}$ && $\tm_1$, $\tm_2$
   && $P^{\pi N N}_{31}$ &=& $\frac 1 6 \left( s^\pi_{111} + 5
s^\pi_{211} \right)$ \\
&& && $P^{\pi N \Delta}_{31}$ &=& $\frac{\sqrt{5}}{6} \left(
s^\pi_{111} - s^\pi_{211} \right)$ \\
&& && $P^{\pi \Delta \Delta}_{31}$ &=& $\frac 1 6 \left( 5
s^\pi_{111} + s^\pi_{211} \right)$ \\
&& && $P^{\eta \Delta \Delta}_{31} $ &=& $s^\eta_1$ \\
\hline
$N_{3/2}$ && $\tm_1$, $\tm_2$
   && $P^{\pi N N}_{13}$ &=& $\frac 1 6 \left( s^\pi_{111} + 5
s^\pi_{211} \right)$ \\
&& && $P^{\pi N \Delta}_{13}$ &=& $\frac{\sqrt{5}}{6} \left(
s^\pi_{111} - s^\pi_{211} \right)$ \\
&& && $P^{\pi \Delta \Delta}_{13}$ &=& $\frac 1 6 \left( 5
s^\pi_{111} + s^\pi_{211} \right)$ \\
&& && $P^{\eta N N}_{13} $ &=& $s^\eta_1$ \\
&& && $F^{\pi \Delta \Delta}_{13}$ &=& $s^\pi_{233}$ \\
\hline
$\Delta_{3/2}$ && $\tm_0$, $\tm_1$, $\tm_2$, $\tm_3$
   && $P^{\pi N N}_{33}$ &=& $\frac{1}{12} \left( 2 s^\pi_{011} + 5
s^\pi_{111} + 5 s^\pi_{211} \right)$ \\
&& && $P^{\pi N \Delta}_{33}$ &=& $\frac{5}{\sqrt{2}} \left(
s^\pi_{011} + s^\pi_{111} - 2 s^\pi_{211} \right)$ \\
&& && $P^{\pi \Delta \Delta}_{33}$ &=& $\frac{1}{15} \left( 5
s^\pi_{011} + 2 s^\pi_{111} + 8 s^\pi_{211} \right)$ \\
&& && $P^{\eta \Delta \Delta}_{33} $ &=& $s^\eta_1$ \\
&& && $F^{\pi \Delta \Delta}_{33}$ &=& $\frac 1 5 \left( s^\pi_{233} +
4 s^\pi_{333} \right)$ \\
&& && $F^{\eta \Delta \Delta}_{33}$ &=& $s^\eta_3$ \\
\hline
$N_{5/2}$ && $\tm_2$, $\tm_3$
   && $P^{\pi \Delta \Delta}_{15}$ &=& $s^\pi_{211}$ \\
&& && $F^{\pi N N}_{15}$ &=& $\frac 1 9 \left( 5 s^\pi_{233} + 4
s^\pi_{333} \right)$ \\
&& && $F^{\pi N \Delta}_{15}$ &=& $\frac 2 9 \sqrt{5} \left(
s^\pi_{233} - s^\pi_{333} \right)$ \\
&& && $F^{\pi \Delta \Delta}_{15}$ &=& $\frac 1 9 \left( 4 s^\pi_{233}
+ 5 s^\pi_{333} \right)$ \\
&& && $F^{\eta N N}_{15}$ &=& $s^\eta_3$ \\
\hline
$\Delta_{5/2}$ && $\tm_1$, $\tm_2$, $\tm_3$, $\tm_4$
   && $P^{\pi \Delta \Delta}_{35}$ &=& $\frac{1}{10} \left( 3
s^\pi_{111} + 7 s^\pi_{211} \right)$ \\
&& && $P^{\eta \Delta \Delta}_{35}$ &=& $s^\eta_1$ \\
&& && $F^{\pi N N}_{35}$ &=& $\frac{1}{126} \left( 10 s^\pi_{233} + 35
s^\pi_{333} + 81 s^\pi_{433} \right)$ \\
&& && $F^{\pi N \Delta}_{35}$ &=& $\frac{1}{126\sqrt{2}} \left( 32
s^\pi_{233} + 49 s^\pi_{333} - 81 s^\pi_{433} \right)$ \\
&& && $F^{\pi \Delta \Delta}_{35}$ &=& $\frac{1}{1260} \left( 512
s^\pi_{233} + 343 s^\pi_{333} + 405 s^\pi_{433} \right)$ \\
&& && $F^{\eta \Delta \Delta}_{35}$ &=& $s^\eta_3$ \\
\hline   
$N_{7/2}$ && $\tm_3$, $\tm_4$
   && $F^{\pi N N}_{17}$ &=& $\frac 1 4 \left( s^\pi_{333} + 3
s^\pi_{433} \right)$ \\
&& && $F^{\pi N \Delta}_{17}$ &=& $\frac{\sqrt{3}}{4} \left(
s^\pi_{333} - s^\pi_{433} \right)$ \\
&& && $F^{\pi \Delta \Delta}_{17}$ &=& $\frac 3 4 \left( 3
s^\pi_{333} + s^\pi_{433} \right)$ \\
&& && $F^{\eta N N}_{17}$ &=& $s^\eta_3$ \\
&& && $H^{\pi \Delta \Delta}_{17}$ &=& $s^\pi_{455}$ \\
\hline
$\Delta_{7/2}$ && $\tm_2$, $\tm_3$, $\tm_4$, $\tm_5$
   && $F^{\pi N N}_{37}$ &=& $\frac{1}{56} \left( 20 s^\pi_{233} + 21
s^\pi_{333} + 15 s^\pi_{433} \right)$ \\
&& && $F^{\pi N \Delta}_{37}$ &=& $\frac 1 7 \sqrt{\frac{15}{2}}
\left( s^\pi_{233} - s^\pi_{433} \right)$ \\
&& && $F^{\pi \Delta \Delta}_{37}$ &=& $\frac 1 7 \left( 3 s^\pi_{233}
+ 4 s^\pi_{433} \right)$ \\
&& && $F^{\eta \Delta \Delta}_{37}$ &=& $s^\eta_3$ \\
&& && $H^{\pi \Delta \Delta}_{37}$ &=& $\frac{1}{25} \left( 7
s^\pi_{455} + 18 s^\pi_{455} \right)$ \\
&& && $H^{\eta \Delta \Delta}_{37}$ &=& $s^\eta_5$ \\
\hline
$N_{9/2}$ && $\tm_4$, $\tm_5$
   && $F^{\pi \Delta \Delta}_{19}$ &=& $s^\pi_{433}$ \\
&& && $H^{\pi N N}_{19}$ &=& $\frac 1 5 \left( 3 s^\pi_{455} + 2
s^\pi_{555} \right)$ \\
&& && $H^{\pi N \Delta}_{19}$ &=& $\frac{\sqrt{6}}{5} \left(
s^\pi_{455} - s^\pi_{555} \right)$ \\
&& && $H^{\pi \Delta \Delta}_{19}$ &=& $\frac 1 5 \left( 2 s^\pi_{455}
+ 3s^\pi_{555} \right)$ \\
&& && $H^{\eta \Delta \Delta}_{19}$ &=& $s^\eta_5$ \\
\hline\hline
\end{tabular}
\end{table}
\begin{table}
\begin{tabular}{lcccccl}
State \mbox{  } && Masses Accommodated \mbox{   } &&
\multicolumn{3}{l}{Partial Wave, $K$-Amplitudes} \\
\hline\hline
$\Delta_{9/2}$ && $\tm_3$, $\tm_4$, $\tm_5$, $\tm_6$
   && $F^{\pi \Delta \Delta}_{39}$ &=& $\frac{1}{20} \left( 9
s^\pi_{333} + 11 s^\pi_{433} \right)$ \\
&& && $F^{\eta \Delta \Delta}_{19}$ &=& $s^\eta_3$ \\
&& && $H^{\pi N N}_{39}$ &=& $\frac{1}{110} \left( 12 s^\pi_{455} + 33
s^\pi_{555} + 65 s^\pi_{655} \right)$ \\
&& && $H^{\pi N \Delta}_{39}$ &=& $\frac{1}{110} \sqrt{\frac{3}{5}}
\left( 32 s^\pi_{455} + 33 s^\pi_{555} - 65 s^\pi_{655} \right)$ \\
&& && $H^{\pi \Delta \Delta}_{39}$ &=& $\frac{1}{550} \left( 256
s^\pi_{455} + 99 s^\pi_{555} + 195 s^\pi_{655} \right)$ \\
&& && $H^{\eta \Delta \Delta}_{39}$ &=& $s^\eta_5$ \\
\hline
$\Delta_{11/2}$ && $\tm_4$, $\tm_5$, $\tm_6$, $\tm_7$
   && $H^{\pi N N}_{3,11}$ &=& $\frac{1}{396} \left( 162 s^\pi_{455} +
143 s^\pi_{555} + 91 s^\pi_{655} \right)$ \\
&& && $H^{\pi N \Delta}_{3,11}$ &=& $\frac{1}{495} \sqrt{\frac{13}{2}}
\left( 81 s^\pi_{455} - 11 s^\pi_{555} - 70 s^\pi_{655} \right)$ \\
&& && $H^{\pi \Delta \Delta}_{3,11}$ &=& $\frac{1}{2475} \left( 1053
s^\pi_{455} + 22 s^\pi_{555} + 1400 s^\pi_{655} \right)$ \\
&& && $H^{\eta \Delta \Delta}_{3,11}$ &=& $s^\eta_5$ \\
&& && $K^{\pi \Delta \Delta}_{3,11}$ &=& $\frac{1}{35} \left( 11
s^\pi_{677} + 24 s^\pi_{777} \right)$ \\
&& && $K^{\eta \Delta \Delta}_{3,11}$ &=& $s^\eta_7$ \\
\hline\hline
\end{tabular}
\end{table}

The principal result of this work is illustrated by the fact that the
quark-picture multiplet structures listed in Table~\ref{qtable} are
found to be completely compatible with those listed in
Tables~\ref{neg} and \ref{pos}.  To be precise, for all values $\ell
\! = \! 0, 1, 2, 3$, one sees that an S multiplet of given $\ell$ contains
only a single mass eigenvalue $m_\ell$, degenerate among all states in
the multiplet up to and including effects of $O(N_c^0)$, the order of
meson-baryon scattering.  This mass eigenvalue is seen to occur for,
{\em and only for}, states that can accommodate a $K$-amplitude with
$K \! = \! \ell$.  Furthermore, an MS or A multiplet of given $\ell$
contains precisely those mass eigenvalues $m_K$ such that $K$ assumes
all values satisfying the triangle rule $\delta (K \ell 1)$.

Thus, for example, each state in the MS $\ell \! = \! 2$ multiplet in
Table~\ref{qtable} assumes one of three possible mass eigenvalues
degenerate to $O(N_c^0)$: $m_1$, $m_2$, or $m_3$. On the other hand,
one can see by referring to Tables~\ref{neg} and \ref{pos} that all
states given in the next-to-last line of Table~\ref{qtable} appear as
poles with $K \!  = \! 1$, 2, or 3, with precisely the right
multiplicities as predicted by Tables~\ref{neg} and \ref{pos}.  Thus
only a single MS $\ell \! = \! 2$ $N_{1/2}$ state occurs, despite the
fact that Tables~\ref{neg} and \ref{pos} allow for two poles with
$N_{1/2}$ quantum numbers ($m_0$ and $m_1$), simply because the $m_0$
pole (from a $K \! = \! 0$ amplitude) does not appear in the MS $\ell
\! = \! 2$ multiplet.  On the other hand, the quark-picture multiplet
contains no $N_{9/2}$ state, while the scattering amplitudes with
$N_{9/2}$ quantum numbers contain no contribution $K \! = \! 1$, 2, or
3, and hence no $m_1$, $m_2$, or $m_3$ pole.

We hasten to add two comments.  The first is that we have demonstrated
only the {\em compatibility\/} of quark-picture SU($2N_F$)$\times$O(3)
and meson-baryon scattering-picture degeneracies among baryon
resonances; we have not proved that the two pictures are the same, but
rather only that they can be assumed simultaneously correct in their
descriptions of the multiplet patterns without contradictions or the
necessity of imposing strong constraints on any configuration mixing
between multiplets.  Indeed, no explicit proof of statements regarding
the degeneracy of mass eigenvalues in the quark-shell picture has been
given thus far (except for previous results for the symmetric $\ell \!
= \! 0$ ground-state multiplet~\cite{Jenk,DJM1,DJM2}, the MS $\ell \!
= \! 1$ excited multiplet~\cite{CCGL1}, and the S $\ell \! = \! 2$
multiplet~\cite{GSS2}), but a general argument (using a hedgehog-based
analysis) appears in Sec.~\ref{gen}.  Second, we have carried out this
demonstration explicitly only up to $\ell \! = \! 3$.  While one might
suspect that the chances of the result not being generic is
exceedingly small, it is still of great value to see a general proof,
as well as to understand the underlying symmetry reason for this
remarkable compatibility.

\section{A General Demonstration} \label{gen}

The previous three sections provide a rather circuitous demonstration
of what appears to be a simple result: Namely, the spectrum of states
for {\em any\/} SU(4)$\times$O(3) multiplet (of either $P$) consists
of a small set of eigenvalues $\left\{m_K\right\}$.  Here, $K$ refers
to $K$-spin value of the meson-baryon (reduced) scattering amplitudes
that can accommodate the $I,J$ quantum numbers of states in the
multiplet possessing the particular eigenvalue $m_K$.  Therefore,
$m_K$ gives the position of a (complex) pole in an amplitude with this
$K$-spin value.  On the other hand, scattering amplitudes with given
$I,J$ quantum numbers do {\em not\/} contain amplitudes of a
particular $K$, and hence have no pole at $m_K$, {\em unless\/} a
state with these $I,J$ values appears in an SU(4)$\times$O(3)
multiplet possessing $m_K$ as one of its mass eigenvalues.

In review, there are two basic pictures for baryon resonances: (i) the
SU(4)$\times$O(3) classification derived from the quark-shell picture
of orbital single-quark excitations about a spin-flavor symmetric core
(the {\em quark\/} or {\em operator\/} picture), and (ii) the
meson-baryon scattering classification derived from symmetry features
shared by all chiral soliton models (the {\em resonance\/} or {\em
scattering\/} picture).  {\em We now show that the underlying reason
for the compatibility of the two pictures is simply that they obey
essentially the same symmetry constraints.}

Two different routes lead to this general result.  To see them,
consider the derivation of Eqs.~(\ref{MPeqn}), or more generally
(\ref{Mmaster}), to describe the symmetry constraints on meson-baryon
scattering.  These relations were first obtained in the context of the
Skyrme model~\cite{HEHW,MP,MK,MM,Mat3}.  In this picture, the quantum
number $K$ has a simple physical interpretation: The Skyrmion is a
chiral soliton at the classical or mean-field level, which at leading
order in large $N_c$ has a hedgehog structure.  Thus it breaks both
the rotational and isospin symmetries separately, but does not break
the grand spin $K$ defined by ${\bf K} \equiv {\bf I}+ {\bf J}$.  The
soliton's intrinsic dynamics (that not associated with collective zero
modes) is governed by a Hamiltonian that commutes with $K$, so that
excitations and scattering states can be labeled by $K$.  The full
relations follow from projecting such states, consisting of an
excitation characterized by $K$ on top of the hedgehog, onto channels
of good $I$ and $J$.

An alternative, more formal, derivation of these results is given in
the appendix of Ref.~\cite{us}.  One uses the fact that matrix
elements in baryon states are fixed by the contracted SU(2$N_F$)
spin-flavor symmetry, which in turn requires that, in order to
contribute at leading order in $1/N_c$, the operator must have $I \! =
\! J$. Writing the meson-baryon scattering amplitude as a matrix
element of a scattering operator in a space of asymptotic baryon $\! +
\!$ meson states, one sees that this operator implies $I_t \! = \!
J_t$, where $t$ indicates the angular momentum and isospin exchanged
in the $t$-channel.  Using standard group-theoretical identities to
cross from the $t$-channel to the $s$-channel gives the
relations~(\ref{Mmaster}).  The $K$ quantum number then emerges as a
summation variable in an identity involving $6j$ coefficients, and
labels the fundamental underlying scattering amplitudes, independent
of $I$ or $J$.

In a similar way we can understand the compatibility discussed above
in terms of either a simple physical picture involving excitations of
hedgehogs, or in terms of a more formal group-theoretical treatment.
We include both treatments here.

\subsection{Physical (Hedgehog) Demonstration}

We begin with the hedgehog-based analysis.  The proof exploits some
simple operator relations. Assume a Hamiltonian $\hat{H}$ possessing
an eigenstate $|\psi_a \rangle$ with eigenvalue $E_a$.  Furthermore,
assume an operator $\hat{\Lambda}$ with the property
\begin{equation}
[\hat{H},\hat{\Lambda}] \, = \, \lambda \hat{\Lambda} \; .
\label{Lambda}
\end{equation}
It is straightforward to see from the commutator that $|\psi_b \rangle
= \hat{\Lambda} |\psi_a \rangle $ either has zero norm or is an
eigenstate of $\hat{H}$ with eigenvalue $E_a + \lambda$.  Thus if $
|\psi_a \rangle $ is the ground state of the system, then
$\hat{\Lambda}$ serves as an excitation operator creating an excited
state.

Now let us apply this to the case of a large $N_c$ version of the
quark-shell model.  This model is based on nonrelativistic quark
degrees of freedom, and has the standard large $N_c$ scaling rules.
The model has rotational and isospin symmetries (with possible small
isospin breaking, which we neglect), so that $I^2$ and $J^2$ are good
quantum numbers of the physical states.  In addition, this model in
its simplest form has no configuration mixing, so that the number of
quarks in a given orbital is also a good quantum number.  Our approach
is to use excitation operators of the form discussed above on this
system, {\it i.e.}, to find some operator $\hat{\Lambda}$ with the
commutator property in Eq.~(\ref{Lambda}) that produces excited states
when acting on the ground state.  However, to leading order in the
$1/N_c$ expansion the ground state is highly degenerate (consisting of
states with $I \!  = \! J \! =$ 1/2, 3/2, \ldots), and thus one can
use any superposition of these states as the ground state.  Our
strategy is to use a hedgehog state as the appropriate superposition.
This approach has the advantage of putting the analysis in close
parallel to the scattering treatment based on the Skyrme model.

The hedgehog state $| h \rangle$ (which has $K \! = \! 0$) is given by
\begin{eqnarray}
| h \rangle \, & = & \frac{1}{\sqrt{N_c!}} \,
\epsilon_{c_1 c_2 \cdots c_{N_c}} \prod_{i=1}^{N_c}
a^\dagger_{h; \, {c_i}} \, | 0 \rangle , \; \; \; {\rm with}
\nonumber \\ a^\dagger_{h; \, c} \, &\equiv & \,
\frac{1}{\sqrt{2}} \left ( a^\dagger_{1/2,-1/2; \, c} - 
a^\dagger_{-1/2,+1/2; \, c} \right ) ,
\end{eqnarray}
where $c$ represents the quark color and $a^\dagger_{m_s,m_i; \, c}$
is the creation operator for a quark in the lowest s-wave orbital with
spin projection $m_s$ and isospin projection $m_i$.  This state has
the (somewhat unfortunate) property that $\langle h |\hat{I}^2| h
\rangle \! = \! O(N_c)$.  Since excitation energies in the
ground-state band are given by $I(I+1)/2 {\cal I}$, where ${\cal I} =
\frac 2 3 (M_\Delta - M_N)^{-1} \! = \! O(N_c)$, one sees that the
$\langle h |\hat{H}| h \rangle - \langle {\rm ground} |\hat{H}| {\rm
ground} \rangle \! = \!  O(N_c^0)$.  Thus one cannot treat the
hedgehog state as degenerate with the ground state at large $N_c$
since the physically interesting quantities---the excitation energies
of other bands---are also $O(N_c^0)$.  It is simple to finesse this
problem: Consider instead eigenstates of the modified Hamiltonian
\begin{equation}
\hat{H}^\prime \equiv \hat{H} - c_2 \hat{I}^2 - c_4 \hat{I}^4 -c_6
\hat{I}^6 - \cdots \; \;,
\label{H'}
\end{equation}
where the coefficients $c_2$, $c_4$, {\it etc.} are fixed to ensure
that all states in the ground-state band are made degenerate.  Note
that each eigenstate of $\hat{H}^\prime$ is an eigenstate of $\hat{H}$
with its eigenvalue shifted by $- c_2 \hat{I}^2 - c_4 \hat{I}^4 -c_6
\hat{I}^6 - \cdots$.  Finding the eigenspectrum of $\hat{H}^\prime$ is
therefore equivalent to finding the eigenspectrum of $\hat{H}$.  By
construction, $| h \rangle$ is degenerate with the ground state with
respect to the operator $\hat{H}^\prime$.

We first consider the case of a single quark excited above the lowest
s-wave orbital.  Recall that in the quark-shell model the number of
excited quarks is well defined.  Now the demonstration that this
degeneracy pattern is compatible with the scattering picture consists
of two parts.  First, we note that {\em if\/} we can show that an
operator $\hat{\Lambda}$ satisfying $[\hat{H}^\prime , \hat{\Lambda}]
= \lambda \hat{\Lambda}$ can be written as a one-body operator that
annihilates a ground-state hedgehog quark and creates a quark in an
excited orbital with good $K$ (where ${\bf K}= {\bf I}+ {\bf J})$,
then this operator acting on the state $| h \rangle$ creates an
excited eigenstate state of the Hamiltonian with well defined $K$ (but
generally not of good $I$ or $J$).  Projecting such a state onto
states of good $I$ and $J$ then produces a set of degenerate states
(mass eigenvalue $m_K$) of distinct $I$ and $J$.  Moreover, in the
large $N_c$ limit the projection onto physical states for this problem
is identical to that done in the Skyrme model for scattering
states~\cite{MP}: In that case one finds scattering states with good
$K$ on top of the hedgehog and uses a semi-classical projection method
with Skyrmion wave functions to obtain channels with good $I$ and $J$.
In the present context, the same projection is needed, only now for
bound states of fixed $K$.  Since the projections in the two cases are
the same, one can reach states of the same quantum numbers in both
cases.  Thus, without further computation, one sees that degenerate
resonances with distinct $I$ and $J$ but labeled by the same intrinsic
$K$ in the scattering case in the Skyrme model are mapped one-to-one
onto degenerate bound states in the quark-shell model.  Note moreover
that, as discussed in Ref.~\cite{MP}, the relations of scattering
amplitudes in the Skyrme model are in fact exact large $N_c$ results,
and thus we see that the degeneracy patterns in large $N_c$ scattering
are also seen in single-quark excitations in the large $N_c$ quark
model.

Before proceeding, it is useful to make a technical remark about the
projection.  In the case of the Skyrmion, a semi-classical projection
was used. In the present context, there is an explicit many-body wave
function, for which the standard Peierls-Yoccoz type
projection~\cite{PYoc} familiar from many-body theory is applicable.
At finite $N_c$, the two types of projection may differ.  However, in
the large $N_c$ limit, the relevant overlap functions become narrow
and the results of explicit projection agree with the semi-classical
result.  The key to this result is that the overlap of a (full
baryonic) hedgehog with a rotated hedgehog goes like the overlap of
single-quark hedgehogs to the $n$th power, where $n$ is in the number
of quarks in the hedgehog configuration.  Since the single-quark
overlap always has a norm of less than or equal to unity and $n \sim
N_c$, one see that as $N_c \rightarrow \infty$ the overlap approaches
zero unless the two hedgehogs are aligned.  This yields narrow overlap
functions and implies the validity of the semi-classical
approximation.

Now to complete this argument, we need to prove the claim that the
excitation operator $\hat{\Lambda}$ satisfying $[\hat{H}^\prime ,
\hat{\Lambda}] = \lambda \hat{\Lambda}$ can, in fact, be written as a
one-body operator that annihilates a ground-state hedgehog quark and
creates quark in an excited orbital with good $K$, and that all
excitations with a single quark in an excited orbital can be projected
from operators of this general form.  To do this, consider the form of
the leading-order Hamiltonian (or more precisely $\hat{H^\prime}$)
from the quark-shell model. From Refs.~\cite{CCGL1,CCGL2} we know that
the leading-order operators that can contribute to single-quark
excitations are of the form of $\openone$, {\boldmath
\mbox{$\ell$}}$\cdot${\bf s}, and $\ell_{ij}^{(2)} g^{i a} G_c^{j a} /
N_c$, where $s^i$ is the matrix element of $\frac 1 2 \sigma^i$ for
the excited quark and ${\bf \ell}$ is its orbital excitation,
$\ell_{ij}^{(2)} \equiv \frac 1 2 \{ \ell^i ,\ell^j \} - \frac 1 3
\delta^{ij} \ell^2$ is the traceless, symmetric tensor obtained solely
from the orbital part of the excited quark (by construction it is a
rank-2 spherical tensor under rotations), $g^{i a}$ is the matrix
element of $\frac 1 2 \sigma^i \otimes \frac 1 2 \tau^a$ for the
excited quark, and $G^{i a}_c$ is the matrix element of $\frac 1 2
\sigma^i \otimes \frac 1 2 \tau^a$ summed over all of the ``core''
quarks in the lowest orbital.  In these expressions repeated indices
are implicitly summed. Note these are one- and two-body operators.
All higher-body operators with $O(N_c^0)$ matrix elements reduce to
these forms using the reduction rules of
Refs.~\cite{DJM2,CCGL1,CCGL2}.  The operators $\openone$ and
{\boldmath \mbox{$\ell$}}$\cdot${\bf s} clearly do not depend on the
state of the core, nor do they change the $K$ value of the excited
quark since both $\openone$ and {\boldmath \mbox{$\ell$}}$\cdot${\bf
s} commute with {\bf j} = {\boldmath \mbox{$\ell$}} + {\bf s}.  The
only one of these operators that mixes core and excited quarks is
$\ell_{i j}^{(2)} g^{i a} G_c^{j a}/N_c$.  To proceed further we must
analyze the effect of this operator in some detail.

The operator $\ell_{i j}^{(2)} g^{i a} G_c^{j a}/N_c$ acts on both the
excited quark and the core.  It is useful to express each component in
terms of pieces that transform under irreducible representations of
$K$.  Accordingly, we write
\begin{eqnarray}
\ell^{(2)}_{ij} g^{i a} G_c^{j a} & = & e^{(0)} G^{(0)} +
 e^{(1,2)}_{i a} G^{(1,2)}_{i a} , \;\;\; {\rm where} \nonumber \\
e^{(0)} & \equiv &  \ell_{i a}^{(2)} g^{i a} , \nonumber \\
 G^{(0)}  & \equiv & \frac 1 3 G_c^{a a} , \nonumber\\
e^{(1,2)}_{i a} & \equiv & \ell_{i j}^{(2)} g^{j a} - 
\frac{1}{3}\delta_{i a} e^{(0)}  , \nonumber \\
 G^{(1,2)}_{i a} & \equiv & G_{c}^{i a} -  \delta_{i a}
G^{(0)} ,
\label{decomp}
\end{eqnarray}
and the superscripts $(0)$ and $(1,2)$ refer to the irreducible
spherical tensor structure under ${\bf K}= {\bf I} + {\bf J}$.  This
decomposition is useful for acting upon the class of states in which
all of the quarks in the core are in a hedgehog configuration.
Suppose one has a state $| \psi \rangle$ in this class.  Writing
components of $G$ in terms of ladder operators, it is straightforward
to see that
\begin{equation}
\frac{1}{N_c} G^{(0)} |\psi \rangle = -\frac 1 4
\frac{n_{\rm core}}{N_c} | \psi \rangle , \; \; \;
 \frac{1}{N_c}G^{(1,2 )}_{i a} |\psi \rangle = O ( n_{core}^{1/2}/N_c ),
\label{g0}
\end{equation}
where $n_{\rm core}$ is the number of quarks in the core; for the case
of a single excited quark $n_{\rm core}=N_c \! - \! 1$. Thus, operating
on a state of this form it is apparent that
\begin{equation}
 \frac{1}{N_c} \ell_{i j}^{(2)} g_{i a} G_c^{j a} |\psi \rangle = 
-\frac 1 4 \frac{N_c-1}{N_c}e^{(0)} |\psi \rangle  +  O(N_c^{-1/2})\; .
\end{equation}
We see that {\em all\/} of the leading-order operators in the $1/N_c$
expansion acting upon states in the class $|\psi \rangle$ are $K \! =
\! 0$ operators acting on the excited quark or core.  This essentially
completes the demonstration since it implies that there are
eigenstates of $\hat{H^\prime}$ with single-quark excitations of good
$K$ acting above a core with all states in the hedgehog; one-body
$\hat{\Lambda}$ transition operators that destroy a core hedgehog
quark and create an excited quark of fixed $K$ [{\it i.e.}, of the
form $\hat{\Lambda} \propto a^\dagger_{\rm exc} (\Delta K \! = \! 0)
a_{\rm core} (\Delta K \! = \! 0)$] satisfy Eq.~(\ref{Lambda}). While
the preceding argument was given explicitly for the case of a single
quark excited outside the core, it is apparent that an analogous
argument holds generally for excitations with $O(N_c^0)$ excited
quarks outside the core.

Finally, we note that the hedgehog picture provides a very simple
physical interpretation of the various spin-flavor symmetries (S, MS,
A).  Again, for simplicity consider first the case of a single quark
outside a core composed entirely of quarks in a spin-flavor hedgehog.
As noted above, these excitations can be labeled by the grand spin $K$
of the excited quark with ${\bf K}_{\rm exc} = {\bf I}_{\rm exc} +
{\bf J}_{\rm exc}$.  Of course, ${\bf J}_{\rm exc}$ has both a spin
part and an orbital part. It is useful to decompose ${\bf K}_{\rm
exc}$ in the following way: ${\bf K}_{\rm exc} =
\mbox{\boldmath{\mbox{$\ell$}}} + {\bf k}$ where ${\bf k} =
{\bf I}_{\rm exc} + {\bf s}$ is purely a spin-flavor construction
independent of the orbital angular momentum.  It is clear that $k$ can
be either 0 or 1, which together saturate the four possible
spin-flavor states of the excited quark.  A $k \! = \! 0$ quark is in
a spin-flavor hedgehog, while a $k \! = \! 1$ quark is orthogonal to
the hedgehog.  It is trivial to see that a state consisting of a
single excited quark in the $k \! = \! 0$ (hedgehog configuration)
above the hedgehog core must be purely symmetric under spin-flavor
since all the quarks are in an identical spin-flavor state (although
radial excitations may be present). Next consider a state
corresponding to a single $k \! = \!  1$ quark outside the hedgehog
core; since the core alone has $K_{\rm core} \! = \!  0$, we may
denote such a state combined with the core as $|k=1 \rangle$ (noting
that the full $K$ of the state still requires the inclusion of orbital
angular momentum $\ell$).  It is straightforward to decompose this
state into a symmetric and a mixed-symmetric piece:
\begin{equation}
|k=1 \rangle = \sqrt{\frac{1}{N_c}}|k=1 \rangle_{\rm S} +
\sqrt{\frac{N_c-1}{N_c}}|k=1 \rangle_{\rm MS} .
\end{equation}
Note that in the large $N_c$ limit such a state becomes purely MS.
Thus for large $N_c$ the following interpretation of the symmetry as
emerges: The S states are those that may be projected from states with
a $k \! = \! 0$ (hedgehog) quark outside the core, while the MS states
are those that may be projected from states with a $k \! = \!  1$
quark outside the core.  A similar analysis can be extended to cases
with more than one quark outside the core.  For example, the A
symmetry arises from two non-hedgehog quarks (antisymmetrized with
respect to each other in spin-flavor) outside a hedgehog core.

The patterns seen in Table~\ref{qtable} may be simply understood in
terms of this picture.  Since the symmetric configurations have $k \!
= \! 0$, one sees immediately that $K \! = \! \ell$, and the symmetric
representations are precisely those one obtains by adding $\ell$ to
the $I \! = \! J$ states in a hedgehog.  The MS states are obtained by
combining the $|k = 1 \rangle$ state with $\ell$, giving total $K$
values that satisfy $\delta (K \ell 1)$.  Similarly, it is easy to
understand the fact the MS and A configurations have the same spectrum
of nonstrange states for large $N_c$.  The reason is simply that A is
obtained by adding two antisymmetric $k \! = \! 1$ quarks outside the
hedgehog core.  However, the antisymmetric combination of two $k \! =
\! 1$ states has a net $k$ of unity since the symmetric combination
must yield $k \! =
\! 0$ and $k \! = \! 2$.  Since only the overall $k$ of the quarks
outside the core matters, A and MS look the same.

\subsection{Group-Theoretical Demonstration}

Now let us turn to the more formal group-theoretical treatment.  The
strategy here is to enumerate explicitly  the particular
representations seen in both the scattering picture and in the
quark-shell model and to show explicitly that they are compatible. To
proceed, we first note that the angular momentum quantum number $L$ is
integral, as are the spin and isospin of the (bosonic) $\pi$ and
$\eta$ mesons.  On the other hand, the baryons involved in the
scattering are of course fermionic, and therefore $R$, and hence $I$
and $J$, are half-integral, while $K$ is integral.  This is
significant because we use that $R$, $I$, and $J$ are nonzero.

We now return to the scattering-picture constraints,
Eqs.~(\ref{MPeqn}), that refer to $\pi$ and $\eta$ scattering.  Note
that the $6j$ symbols of Eq.~(\ref{MPeqn})---considering only quantum
numbers of the initial state---imply the four triangle rules $\delta
(L R J)$, $\delta (R \, 1 I)$, $\delta (L \, 1 K)$, and $\delta (I J
K)$.  First we ask whether amplitudes for $\eta$ scattering can ever
possess quantum numbers violating any of these triangle rules.  The
constraints in the second of (\ref{MPeqn}) consist of $\delta (L R
J)$, but also $I \! = \! R$ (because the $\eta$ carries zero isospin)
and $K \! = \! L$.  Now, $I \! = \! R$ always satisfies $\delta (R \,
1 I)$ because $R \neq 0$.  On the other hand, $K \! = \! L$ always
satisfies $\delta (L \, 1 K)$ unless $K \! = \! L \! = \! 0$.  It
follows that $\eta$-baryon scattering allows different amplitudes than
in $\pi$-baryon scattering only if $K \! = \! L \! = \! 0$, $I \! = \!
R$, and hence [because of $\delta (L R J)]$, $I \! = \! J$.  On the
other hand, it is clear that $\pi$-baryon scattering allows many
amplitudes with quantum numbers not allowed in $\eta$ scattering.

Now, given the $I,J$ quantum numbers of a scattering channel, what
values of $K$ appear?  From Tables~\ref{neg} and \ref{pos}, it is
clear for all channels considered that the full set $\delta (I J K)$,
$|I\!-\!J| \! \leq \! K \! \leq \! I\!+\!J$, appears.  However,
Eqs.~(\ref{MPeqn}) give additional constraints on $K$ that must be
taken into account: One must check that values for $R$ and $L$ exist
[satisfying $\delta (L R J)$, $\delta (R \, 1 I)$, and $\delta (L \, 1
K)$ for the $\pi$ case, or $I \! = \! R$ and $K \! = \! L$ for the
$\eta$ case] that allow the full range of $K$ for arbitrary
half-integral $I$ and $J$.  $P$ conservation places no constraints on
the group theory except that it must be conserved throughout any
process, and $P \! = \! (-1)^{L+1}$ for the resonant state since the
scattered meson is pseudoscalar; therefore, we wish to prove the
parity blindness of the scattering picture by showing that there exist
an even and an odd value of $L$ both satisfying the above constraints.

Consider the value $R \! = \! I$, which always satisfies $\delta (R \,
1 I)$ since $I\!\neq\!0$.  Further, let $L \! = \! K$, which always
satisfies $\delta (L \, 1 K)$ unless $K \! = \! 0$ (a case we handle
in a moment).  Then the final $6j$ symbol $\delta (L R J)$ becomes the
same as $\delta (K I J)$, which is satisfied by assumption.  This
means that a scattering amplitude of any $I,J$ contains a $\pi$
scattering amplitude $s^\pi_{KLL}$, where $K$ is any nonzero value
satisfying $\delta (I J K)$, for which the scattered baryon has $R \!
= \! I$ and for which $L \! = \! K$.  Further, $\delta (L \, 1 K)$ for
$K\!\neq\!0$ is satisfied by three consecutive values of $L$, namely,
$K\!-\!1$, $K$, and $K\!+\!1$.  Since $R$ and $J$ are nonzero, at
least two consecutive values of $L$ satisfy $\delta (L R J)$, one of
which, we have seen, is $L \! = \!  K$.  Thus, for $K\!\neq\!0$, this
proves that at least two consecutive values of $L$ satisfy all
constraints, allowing both parities~\cite{exception}.  As for $K \! =
\! 0$, the $\pi$ scattering constraint $\delta (L \, 1 K)$ is
satisfied solely by $L \!  = \! 1$, while the $\eta$ scattering
constraint $\delta_{KL}$ is satisfied solely by $L \! = \! 0$, giving
the required two consecutive values of $L$ satisfying all constraints.
We conclude that {\em a scattering amplitude with any choice of
$I,J,P$, contains a reduced amplitude with every value $K$ that
satisfies $\delta (I J K)$.}

Now consider the spin-flavor multiplets S, MS, and A discussed in the
previous two sections, and in particular the nonstrange $N$ and
$\Delta$ multiplets listed in Table~\ref{qtable}.  We begin with the S
multiplet, anticipating that the proof for MS and A requires little
additional effort.  The S multiplet for $\ell \! = \! 0$ consists of
states with $I \! = \! S$, where $S$ again is the total quark spin,
and thus the total baryon spin is given by values allowed by {\bf J} =
{\bf S} + {\boldmath \mbox{$\ell$}}, which of course imposes the
triangle rule $\delta (I J \ell)$.  Any such triad of values
$I,J,\ell$ (plus $P$) gives a unique state in a particular $S$
multiplet, as may be seen from the decomposition of S in
Eq.~(\ref{Sdecomp}) and the nondegeneracy of resultant multiplet
quantum numbers under angular momentum addition.  The claim of
compatibility of the two pictures for the S multiplet consists of the
following necessary and sufficient conditions:
\begin{itemize}
\item
If a state with given values of $I,J,P$ appears in an S multiplet with
a given allowed value of $\ell$, then the scattering amplitude with
$I,J,P$ quantum numbers contains a reduced amplitude with $K \! = \!
\ell$.

\item
If a scattering amplitude with given values of $I,J,P$ contains an
amplitude with a particular allowed value of $K$, then a state with
these quantum numbers appears in an S multiplet with $\ell \! = \! K$.
\end{itemize}

The first statement is proved simply by noting that every value of
$\ell$ satisfying $\delta (I J \ell)$ gives precisely one state
appearing in a particular S multiplet for each value of $P$, while we
have just proved that each value of $K$ satisfying $\delta (I J K)$
(including $K \! = \! \ell$) appears in an amplitude with the given
$I,J,P$ quantum numbers for at least one $\pi$ or $\eta$ scattering
process.

The second statement is proved by noting that an amplitude with a
given value of $K$ occurs if it satisfies $\delta (I J K)$, regardless
of $P$.  Letting $K \! = \! \ell$, we see that $\delta (I J \ell)$ is
also satisfied.  Including $P$, this specifies (as we have seen) a
unique state in an S multiplet.  The proof for S multiplets is thus
complete.

Now we return to the MS and A multiplets.  Our task is simplified by
the fact that we are interested only in $N$ and $\Delta$ states, which
appear (Table~\ref{qtable}) with precisely the same multiplicities in
these two representations.  Moreover, we have seen [Eq.~(\ref{Seff})]
that MS or A multiplets with given $\ell$ contain the same $N$ and
$\Delta$ multiplets as given by a effective collection of S
multiplets, with values $\ell_{\rm eff}$ being all those that satisfy
$\delta (\ell_{\rm eff}, \, \ell, \, \Delta \! = \! 1)$.  The proof is
thus reduced to one of showing the compatibility between the
scattering picture and the quark-shell picture states for a collection
of effective S multiplets with $\ell$ values given by $\ell_{\rm
eff}$.  But this is accomplished simply by recourse to the proof we
have just completed: Compatibility is achieved for these effective S
multiplets by placing a pole in each amplitude satisfying $K \! = \!
\ell_{\rm eff}$.  The proof for MS and A multiplets is thus complete.

We arrive at a simple result.  {\em Compatibility between the
meson-baryon scattering and quark-shell pictures is achieved in the
following way: If the orbital excitation of the excited quark in the
quark-shell SU(4)$\times$O(3) multiplet is given by $\ell$, then for
states in S multiplets there exists a scattering partial wave with the
quantum numbers of the given state containing a $K$-amplitude with $K
\! = \! \ell$ , and for states in MS or A multiplets there exist
scattering amplitudes containing $K$-amplitudes for each value of $K$
satisfying $\delta (K \ell 1)$.  Furthermore, reduced amplitudes with
these values of $K$ do not appear in scattering amplitudes with
quantum numbers corresponding to states not appearing in the given
SU(4)$\times$O(3) multiplet.}  The fact that the masses and widths of
resonances may be labeled by $K$ and correspond to poles in the
$K$-amplitudes completes the demonstration of compatibility.
 
It is especially interesting to note that full compatibility would not
occur without the possibility of resonance creation both via
scattering with $\pi$'s and $\eta$'s.  On the other hand, while it may
be tempting to believe that the required inclusion of the pion is an
indication of chiral symmetry, the parity blindness of the group
theory implies that using light positive-parity mesons with $I \! = \!
0$ and 1 would have worked just as well.

\section{Discussion} \label{discuss}

We have compared two popular (and often competing) pictures for baryon
resonances in the large $N_c$ limit.  Both have physical probative
value since both explore different aspects of hadronic physics.  On
one hand, the quark-shell picture certainly is the most successful
model in existence for predicting much of the observed spectrum of
baryon states, but there is no known good reason that quarks should
play the role of the defining degrees of freedom in the highly
confining regime of the hadronic environment.  On the other hand,
baryon resonances truly are observed as ``bumps'' (or even more
obscure features) in meson-baryon scattering amplitudes, but the
scattering picture gives no hints as to which amplitudes should even
possess resonances, much less where the pole positions should lie.

The fact that each of the two pictures respects the spin-flavor
multiplet structure of the other is certainly good news for
phenomenology.  It implies the existence of a sensible way to perform
the $1/N_c$ expansion for baryon resonances.  Namely, these broad
states [widths of $O(N_c^0)$] can be treated in a Hamiltonian
formalism, just as was done so successfully for the ground-state band,
and this treatment respects the symmetries ($I_t \! = \! J_t$)
implicit in the states' origins as resonances in scattering
amplitudes.

However, we caution the reader that what has been presented here is no
more than an existence proof of a unified $1/N_c$ expansion; the full
treatment awaits an analysis of allowed $1/N_c$ effects~\cite{future}.

For example, consider the status of configuration mixing in light of
our results.  The fact that precisely the same nonstrange multiplets
arise for MS and A flavor representations in large $N_c$ implies that
each of the states in one of these multiplets is, in principle, free
to mix with its partner in the other multiplet.  Detailed quark models
tend to disfavor the appearance of A multiplets, chiefly because they
require the excitation of two quarks (see Fig.~\ref{young}), which is
usually suppressed ({\it e.g.}, in one-gluon exchange models).
However, the combinatorics of the $N_c \! - \! 2$ core quarks negates
this suppression in large $N_c$.  Moreover, if $1/N_c$ effects are
properly included, they must eventually be sensitive to the
differences between the MS and A multiplets when $N_c \! \le \! 7$.

Configuration mixing need not be limited to entire spin-flavor
multiplets.  For example, the $K \! = \! 2$ pole $\tilde{m}_2$
appearing in the S multiplet $2^+$ is precisely the same one as
appearing in the $K \! = \! 2$ MS (or A) multiplet $1^+$, $2^+$, or
$3^+$.  Detailed dynamics may suppress some of these spin-flavor
multiplets, but some degree of configuration mixing may still be
expected at large $N_c$, such as between $N_{3/2} ({\bf S}, 2^+)$ and
$N_{3/2} ({\bf MS}, 2^+)$.  Indeed, such mixing is common even in the
$N_c \! = \! 3$ quark model~\cite{FC}.

As suggested above, a full phenomenological analysis of the higher
baryon multiplets must await an accounting of the $1/N_c$ corrections.
From the theoretical side, this is clearly necessary not only because
operators corresponding to important physical effects might first
arise at $O(1/N_c)$, but because the resonance spectrum itself is
different for $N_c \! > \! 3$, and such states must be consistently
decoupled before one has a fully usable formalism for studying the
$N_c \!  = \! 3$ case.  From the experimental side, a quick glance at
the status of the higher baryon states~\cite{PDG} indicates a number
of resonances of dubious existence; assuming the verity of phantom
states (or missing relevant states) could easily lead to a false
interpretation about which underlying dynamical effects are most
important.

In short, the $1/N_c$ expansion for baryon resonances should not be
viewed as the final arbiter of competing dynamical pictures, but
rather as a baseline of symmetry constraints, upon which new ideas for
dynamics can be studied.

\acknowledgments
T.D.C.\ acknowledges the support of the U.S.~Department of Energy
through grant DE-FG02-93ER-40762.  R.F.L.\ acknowledges support from
the National Science Foundation under Grant No.\ PHY-0140362, and
thanks S. Capstick for valuable discussions.


\begin{thebibliography}{99}
\bibitem{exception}
There is one exception to this proof, which is quite rare, but is
dispensed with here for completeness: It is possible for a $6j$ symbol
to vanish even if all relevant triangle rules are satisfied.  This can
happen for the $\pi$ scattering case with $I\!=\!R$, $L\!=\!K\!>\!0$
discussed in the text; however, in this case the $\eta$ amplitude with
the same quantum numbers is always allowed [see Eq.~(\ref{MPeqn})],
and contains the desired amplitude $s^\eta_K$.  The only such
``exceptional zero'' occuring in this analysis is the case
$I\!=\!R\!=\!3/2$, $L\!=K\!=\!3$ $J\!=\!7/2$, the $F^{\pi \Delta
\Delta}_{37}$ amplitude for $\Delta_{7/2}$ in Table~\ref{pos}, where
the $K\!=\!3$ amplitude is seen to be absent.  Indeed, this is the
only such case up to at least $K\!=\!50$!  The $6j$ symbols
corresponding to $L\!=\!K\!\pm\!1$ are always nonzero if they satisfy
all relevant triangle rules, as can be shown from an explicit
expression~\cite{edmonds}.
\end{thebibliography}
\end{document}